\documentclass[twocolumn,prd,superscriptaddress,preprintnumbers,amsmath,amssymb,nofootinbib]{revtex4}

\usepackage{amsmath}
\usepackage{amssymb}
\usepackage{slashed}
\usepackage{graphicx}
\usepackage{graphics}
\usepackage{epsfig}
\usepackage{color}
\usepackage{longtable}
\usepackage{tabularx}
\usepackage{hyperref}

\newcommand{\be}{\begin{equation}}
\newcommand{\ee}{\end{equation}}
\newcommand{\bea}{\begin{eqnarray}}
\newcommand{\eea}{\end{eqnarray}}
\newcommand{\Eqref}[1]{Eq.~\eqref{#1}}

\begin{document}
\title{The Planck scale, the Higgs mass and scalar dark matter}

\author{Astrid Eichhorn}
\email{aeichhorn@perimeterinstitute.ca}
\affiliation{\mbox{\it Perimeter Institute for Theoretical Physics, 31 Caroline Street N, Waterloo, N2L 2Y5, Ontario, Canada}}

\author{Michael M. Scherer}
\email{scherer@thphys.uni-heidelberg.de}
\affiliation{\mbox{\it 	Institut f\"ur Theoretische Physik,
Universit\"at Heidelberg,}
\mbox{\it D-69120 Heidelberg, Germany}
}

\begin{abstract} 
This study is inspired by a scenario, in which the Standard Model, enhanced by an additional dark matter scalar, could be extended up to the Planck scale, while accommodating the low measured value of the Higgs mass. To that end, we study a toy model for a gauge singlet dark matter scalar coupled to the Higgs-top-quark sector of the Standard Model. 
Using functional methods to derive Renormalization Group flow equations in that model, we examine several choices for the ultraviolet, bare potential in the Higgs-dark-matter sector. 
Our results indicate that the dark matter scalar can decrease the lower bound on the Higgs mass in the Standard Model.
We then use the fact that higher-order couplings which are driven to tiny values by the Renormalization Group flow towards low energies can easily be of order one at the ultraviolet cutoff scale. 
Our study indicates that the inclusion of these couplings can significantly increase the ultraviolet cutoff scale and therefore the range of validity of the model while yielding a low value for the Higgs mass in the infrared.
This is achieved within a setting where the dark matter scalar accounts for the complete dark matter relic density in our universe.

\end{abstract}

\maketitle

%%%%%%%%%%%%%%%%%%%%%%%%%%%%%%%%

\section{Higgs mass bounds and dark matter -- two unrelated problems?}

The discovery of the Higgs particle at the LHC \cite{Aad:2012tfa,Chatrchyan:2012ufa} has far-reaching implications for fundamental physics: The Standard Model is a complete low-energy effective theory,  and no new physics at scales close to that probed by the LHC is required for its consistency (which of course does not preclude the possibility that new physics will be found). This opens the door to an exciting possibility, namely that the Standard Model (plus right-handed neutrinos)  describes the fundamental constituents of our universe and their dynamics up to the Planck scale, where quantum gravity effects become important. Thus our knowledge of fundamental particles and their dynamics could hold over a huge range of scales. In particular the absence of new physics on intermediate scales would imply that the value of the Higgs mass depends directly on the effective dynamics at the Planck scale.

This scenario faces at least two crucial challenges:
Firstly, the measured value of the Higgs mass indicates that the maximum value for the ultraviolet (UV) cutoff scale of the Standard Model lies below the Planck mass
\cite{Maiani:1977cg,Altarelli:1994rb,Casas:1994qy,Casas:1996aq,Schrempp:1996fb, Isidori:2001bm, Isidori:2007vm, Ellis:2009tp, EliasMiro:2011aa, Degrassi:2012ry,Bezrukov:2012sa, Buttazzo:2013uya,Kobakhidze:2014xda}, suggesting that we miss degrees of freedom, which become important at that mass scale.
Secondly, cosmological and astrophysical data provide evidence that the Standard Model cannot be the complete story of particle physics, as it does not include dark matter. Assuming that the solution to the dark-matter puzzle indeed lies in particle physics and not in a modification of gravity implies that we miss particle physics degrees of freedom.

It has been proposed, see \cite{Silveira:1985rk,McDonald:1993ex,Bento:2000ah,Burgess:2000yq,Bento:2001yk,McDonald:2001vt,Davoudiasl:2004be,
Patt:2006fw,
O'Connell:2006wi,Barger:2007im,He:2007tt,Barger:2008jx,He:2008qm}, that the second problem could find a solution in the introduction of a single scalar field that serves as a stable dark matter candidate. 
Most interestingly, the -- at a first glance completely unrelated --  
question whether the range of validity of the Standard Model can be extended up to the Planck scale is affected by that same field:
As has been studied in 
\cite{Gonderinger:2009jp}, the additional scalar could lower the minimal allowed value for the Higgs mass at a given UV cutoff scale, see also \cite{Clark:2009dc,Lerner:2009xg,Gonderinger:2012rd,Chen:2012faa}. 
One might hope that a very minimal extension of the Standard Model, adding only a gauge singlet scalar field, could thus solve these two major challenges.
Several studies \cite{Gonderinger:2009jp,Khoze:2014xha} have addressed this question, employing a restricted choice for the ultraviolet potential.

Before we embark on an extended study of this problem, let us recall how the lower bound on the Higgs mass arises in the context of the Standard Model: It is the effect of
 a struggle between bosonic and fermionic fluctuations: Following the Renormalization Group (RG) flow towards the infrared, bosonic fluctuations decrease the value of the quartic Higgs self-coupling. On the other hand, fermionic fluctuations, in particular of the top quark, yield an increase of the low-energy value of that coupling.  Since the Higgs mass is directly proportional to the Higgs self-coupling, a dominance of top-quark fluctuations yields a large value of the Higgs mass in the infrared. The minimal value for the Higgs mass, as a function of the UV cutoff $\Lambda$, arises if the microscopic value of the Higgs self-coupling is fixed to the minimal value that is compatible with an ultraviolet stable potential, i.e., a vanishing value. In the case of a UV cutoff at $\Lambda = M_{\rm Planck}$, the lower bound on the low-energy value of the Higgs mass, given a top mass of $m_{\rm top} = 173 \, \rm GeV$, turns out to be higher than the experimentally determined value. 

%%%%%%%%%%%%%%%%%%%%%%%%%%%%%%%%

\subsection{The Standard Model as a low-energy effective field theory}

Here we will suggest a solution to this problem, which exploits the fact that the Standard Model is a low-energy effective theory. This implies of course that at UV scales the Standard Model  does not only contain perturbatively renormalizable (i.e., marginal and relevant), but also canonically irrelevant, i.e., perturbatively non-renormalizable couplings. Such operators, e.g., a $h^6$ interaction for the Higgs field $h$, do not play a role at low energies, as they die out during the Renormalization Group flow towards the infrared (IR). Conversely, they can easily take values $\mathcal{O}(1)$ in the UV and thereby completely alter the RG flow of the effective potential at high scales. The inclusion of such higher-order couplings could have a crucial influence on the question of a lower Higgs-mass bound.

At a first glance, one might object that the validity of an effective-field theory framework for the Standard Model must imply that low-energy observables cannot depend on perturbatively non-renormalizable operators, as these must be suppressed. Here it is crucial to realize, see \cite{Gies:2013fua}, that the question whether the low-energy value of the Higgs mass is compatible with a stable electroweak vacuum, is \emph{not} a low-energy question: It depends on the high-energy values of the couplings in the effective potential, simply because the lower bound on the Higgs mass is a function of $\Lambda$, and thus sensitive to physics at $\Lambda$.

Previous studies  of the RG flow in the Higgs-dark-matter sector ignored these couplings. Since there is no symmetry principle to set those couplings to zero, we generically expect them to be non zero at the UV scale. Thus previous studies employed a particular, fine-tuned choice for these couplings.

In general, the values of those couplings at a UV scale, e.g., the Planck scale, are determined by the ultraviolet completion of the theory, which should also involve gravity. An example of a candidate for a UV completion, which is consistent with the Standard Model data, and would predict particular values for those couplings, is given by asymptotically safe gravity \cite{Reuter:2012id,Zanusso:2009bs,Shaposhnikov:2009pv, Dona:2013qba}. More generally one should expect the following situation: Similarly to QED, where integrating out fermionic fluctuations leads to the Euler-Heisenberg Lagrangian for the low energy photons, the microscopic degrees of freedom constituting the UV completion of the Standard Model could generate all possible operators that are compatible with the symmetries of the low-energy description. Accordingly, operators such as $h^6$ will be present at high scales. In our analysis we will assume that we can employ the effective-field theory framework, i.e., ignore the microscopic degrees of freedom such as quantum gravity fluctuations, in the vicinity of the cutoff scale $\Lambda$, where the induced higher-order couplings are nonzero. Towards the infrared, all but the marginal and relevant couplings will be driven to near-zero values by the Renormalization Group flow. Accordingly the low-energy theory only contains those couplings known from the Standard Model.

In the following, we will not assume anything about the ultraviolet completion. Correspondingly, the values of irrelevant couplings at the UV cutoff scale will be free parameters. We will investigate whether it is possible to find  values for these parameters, such that the vacuum of the Standard Model, augmented by the additional dark matter scalar, could be stable at those high scales, and the measured value for the dark matter relic density and the Higgs mass is reproduced. Invoking a toy model of the Standard Model, we propose a mechanism to achieve these objectives.
Our toy model includes a $\mathbb{Z}_2$-symmetric potential for a scalar, to be identified with the Higgs. This choice is motivated by the fact that the spontaneous breaking of this discrete symmetry will not feature any Goldstone bosons. Similarly, there are no massless Goldstone modes in the Standard Model, as the would-be-Goldstone modes constitute the massive longitudinal polarizations of the W and Z bosons. This toy model without the dark matter scalar has been studied in a similar context by various methods \cite{Holland:2003jr,Holland:2004sd,Fodor:2007fn,Gies:2013fua}.

Further, our model features a fermionic degree of freedom that increases the low-energy value of the Higgs mass through a Yukawa-coupling to the scalar, just as the top quark does for the Standard Model: As top quark fluctuations yield a negative contribution to the beta function of the Higgs self-coupling, they increase the minimal value for the Higgs mass in the infrared.
In our analysis, we will focus on values of the scalar and the fermionic masses that equal the Standard Model masses for the Higgs and the top quark.

Additionally, we include a further scalar $S$, the dark matter particle, which couples to the Standard Model through the Higgs portal, i.e., a quartic $S^2\,  h^2$ interaction.

To evaluate the RG flow within this toy model, we employ the functional Renormalization Group to derive the beta functions for all running couplings $g(k)$, where
\be
\beta_g = k \partial_k\, g(k),
\ee
with a momentum scale $k$. For details on the functional Renormalization Group and our expressions for the $\beta$~functions, see app.~\ref{FRG}. A clear advantage of this method lies in its applicability beyond the range of validity of perturbation theory.

The Renormalization Group flow will then allow us to find a map from the values of the couplings at a UV cutoff scale to the infrared values of the couplings, which are accessible to measurements. Using this map, and requiring positivity and boundedness of the highest-order non-vanishing coupling at the UV cutoff scale, enables us to determine the value of that scale.

A related study has been performed in \cite{Gies:2013fua}, where the RG flow of a toy model of the Higgs sector was studied, taking into account higher-order couplings. In our work, the central assumption is that the Standard Model completed by the Higgs portal to dark matter is the physically relevant model. Therefore the dark matter scalar constitutes a relevant degree of freedom that must be included in the study. 

The effects of higher-order couplings in the Higgs sector have also been addressed previously within a perturbative framework in \cite{Burgess:2001tj, Datta:1996ni, Grzadkowski:2001vb}, where the focus was on scenarios where the presence of higher-order couplings lowers the scale of New Physics.

%%%%%%%%%%%%%%%%%%%%%%%%%%%%%%%%

\section{Higgs-dark-matter model}

We consider a model defined by the action
\bea
S_\mathrm{UV} &=& \int d^4 x \Big\{\bar\psi i \slashed{\partial} \psi
+\frac{1}{2}(\partial_\mu h)^2+\frac{1}{2}(\partial_\mu S)^2\nonumber\\
&&\quad\quad\quad\quad\quad\quad+i \bar y\,  h\bar\psi\psi+\bar V(h,S)\Big\}\,,
\label{effpot}
\eea
where $\psi$ is a Dirac fermion parametrizing the top quark. The real scalar boson $h$ stands for the radial component of the Higgs particle and couples to the top quark by the Yukawa interaction $\bar y$. This part of the action is invariant under $\mathbb{Z}_2$-transformations\footnote{The $\mathbb{Z}_2$ transformation acts as follows:
$
h \rightarrow -h, \quad  \bar{\psi}  \rightarrow  \bar{\psi} e^{i \frac{\pi}{2}\gamma_5}, \quad
\psi \rightarrow e^{i \frac{\pi}{2}\gamma_5} \psi, \quad S \rightarrow S. \nonumber
$
Note that the action on $S$ is trivial.
}. To include dark matter, we introduce a further gauge-singlet scalar field $S$ invariant under a separate $\mathbb{Z}_2$-transformation which leaves invariant all other fields. This discrete symmetry ensures the stability of the scalar and makes it a candidate for dark matter. Self-interactions of the bosonic degrees of freedom and mutual interactions among them are parametrized by the potential $\bar V(h,S)$.

As a crucial difference to earlier investigations of the Higgs portal \cite{Gonderinger:2009jp,Lebedev:2012zw,Haba:2013lga} we now consider a more general effective potential, and extend the expansion to include couplings of higher order in the fields:

\bea
\bar V(h, S)&=& \frac{\bar m_h^2}{2} h^2+ \frac{\bar m_S^2}{2} S^2+\frac{\bar\lambda_{20}}{8}h^4+\frac{\bar\lambda_{02}}{8}S^4\\
&&+\frac{\bar\lambda_{11}}{4}h^2S^2+\frac{\bar\lambda_{30}}{48}h^6+\frac{\bar\lambda_{03}}{48}S^6+...\nonumber
\eea
All couplings depend on a Renormalization Group scale, thus the effective potential is RG scale dependent. In the following, we will employ a dimensionless version of all couplings, obtained by multiplying the couplings with appropriate powers of a momentum scale $k$, i.e., $\lambda_{11} = \bar{\lambda}_{11}, \lambda_{30} = \bar{\lambda}_{30} k^2$ etc.

Note that power-counting suggests that the higher-order, irrelevant couplings $\lambda_{30}, \lambda_{03},...$ will be very small at low energies and the infrared physics is dominated completely by the relevant and marginal couplings. Our numerical evaluation of RG trajectories will explicitly confirm this expectation in agreement with standard effective-field theory arguments. Conversely, these couplings will generically be $\mathcal{O}(1)$ at the UV cutoff scale. Of course, the same argument applies to higher couplings, of e.g., the $h^8$ operator. In our study, we do not aim for a comprehensive investigation of the space of bare potentials. We will show, using the example featuring only a subset of possible higher-order couplings, that their inclusion can provide a mechanism to obtain compatibility between a high value for the UV cutoff scale, and a low value for the Higgs mass in the infrared.

Here we aim for a minimal explanation of dark matter in the sense that the we try to explain the entire dark matter relic density by postulating only one additional degree of freedom, namely the scalar $S$. This requirement imposes a relation between $\lambda_{11}$ and $m_S$, as the relic density depends on the annihilation cross-section of dark matter into other particles, i.e., Standard Model fields. Since the cross-section is a function of both $\lambda_{11}$ and $m_S$, a relation between the two follows from requiring the cross-section to take the correct value to reproduce the dark matter relic density, see \cite{McDonald:1993ex,Burgess:2000yq, Cline:2013gha}.
Taking into account additional Higgs-dark-matter couplings, such as, e.g.,  $S^2 (h^2)^2$, which could also be present at the UV scale could of course change the cross-section. To decide, whether these couplings play a role for dark-matter-annihilation into Standard Model particles, we need to consider the energy scale at which dark matter freezes out. As this is much  lower than the ultraviolet cutoff scale, effective-field-theory reasoning therefore suggests that the presence of higher-order couplings at the UV cutoff scale $\Lambda$ does not affect the annihilation cross-section that determines the dark matter relic density. Accordingly the analysis in \cite{McDonald:1993ex,Burgess:2000yq, Cline:2013gha}  also holds in our case.

\subsection{The effect of dark matter -- first steps}
As a first step we re-examine the system described by \Eqref{effpot}, setting all higher-order couplings to zero. The system is then parametrized by $\lambda_{20}(k), \lambda_{11}(k), \lambda_{02}(k), y(k)$ and $m_{h}(k), m_{S}(k)$. 
While the infrared values of the coupling $\lambda_{11}$ and the mass $m_S$ are related by the observed value of the dark matter relic density, their RG flow and their values at higher scales are not related, since the condition on the scattering cross-section only has to hold at the freeze-out scale.
As discussed in \cite{Cline:2013gha}, the complicated relationship between $m_{S}$ and $\lambda_{11}$ goes over into a simple linear relationship between the logarithms of the two quantities at $m_{S}> m_h/2$. Using\footnote{Note that the difference to the equation given in \cite{Cline:2013gha} arises from a factor of 2 in the definitions of the coupling. While the relation in \cite{Cline:2013gha} is obtained in the context of the Standard Model, we will apply it for our toy model here.} 
\be\label{eq:relic}
{\rm log}_{10} \left(\frac{\lambda_{11}}{2}\right) = -3.63 + 1.04 \, {\rm log}_{10}(m_{S}/ \rm GeV),
\ee
we note that a small value of $\lambda_{11} = 1/10$ already translates into $m_{S} = 173.5 \, \rm GeV$, for which clearly $m_{S}> m_h/2$. Accordingly we can use the above approximation to set the relationship between $m_{S}$ and $\lambda_{11}$ at the infrared scale as we will only be interested in the case $\lambda_{11} \geq 1/10$.

We now set the Higgs vacuum expectation value to $vev=246 \, \rm GeV$ at the infrared scale $k_{\mathrm{IR}}$. Furthermore we set $y(k_{\rm IR}) = 173/246$, thus yielding a top-mass of $m_{\rm top} = 173 \, \rm GeV$. We then construct RG trajectories that yield different values of $\lambda_{20}, \lambda_{11}$ and $\lambda_{02}$ in the infrared as a function of the UV cutoff scale. At the UV cutoff, $\lambda_{20}=0$ yields the lower bound on the Higgs mass in the infrared, whereas $\lambda_{20}(k_{\rm UV})= \lambda_{20\, \rm max}$ yields the upper bounds, related to the triviality problem.
 Note that with our defintions the coupling $\lambda_{20}$ is related to the Higgs mass by $m_H=\sqrt{\lambda_{20}}\cdot vev$. The requirement of a stable potential implies \cite{Burgess:2000yq}:
\bea
&{}& \lambda_{20}(k)>0, \, \, \lambda_{02}(k)>0,\nonumber\\
&{}& \lambda_{11}^2(k) < \lambda_{20}(k) \lambda_{02}(k) \mbox{  for $\lambda_{11}(k)<0$.}
\eea
Furthermore we demand that none of the couplings hits a Landau pole. We implement this condition by demanding that none of the couplings exceeds a given value $\lambda_{\rm max}$, e.g., 10. In the functional RG approach the couplings are not bound by the perturbative regime and we can study couplings with larger values than those typically investigated in a perturbative setup.

\begin{figure}[!here]
\includegraphics[width=\linewidth]{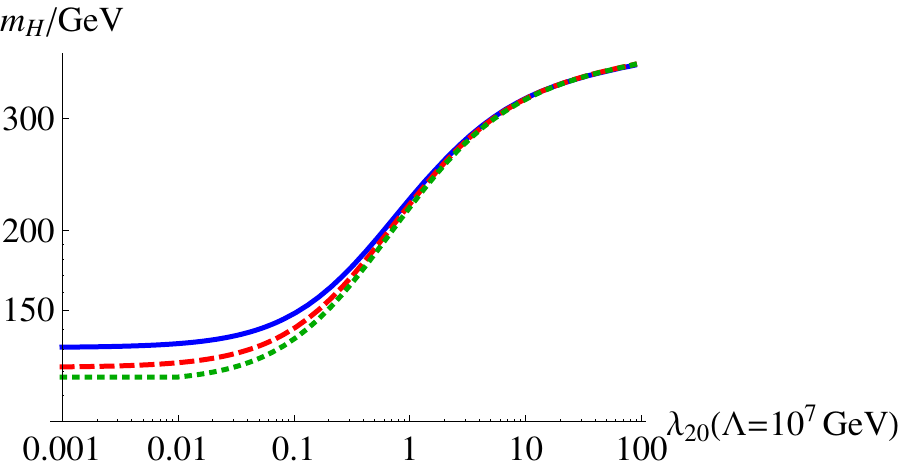}
\caption{\label{CutoffVsMh} We plot the lower bound on the Higgs mass as a function of the UV cutoff scale $\Lambda$. The thick blue line shows the case without dark matter, whereas the other cases include dark matter, with $\lambda_{11}(k_{\rm IR}) = 1$ (green dotted line) and $\lambda_{11}(k_{\rm IR})=0.8$ (red dashed line).}
\end{figure}

\begin{figure}[!here]
\includegraphics[width=0.8\linewidth]{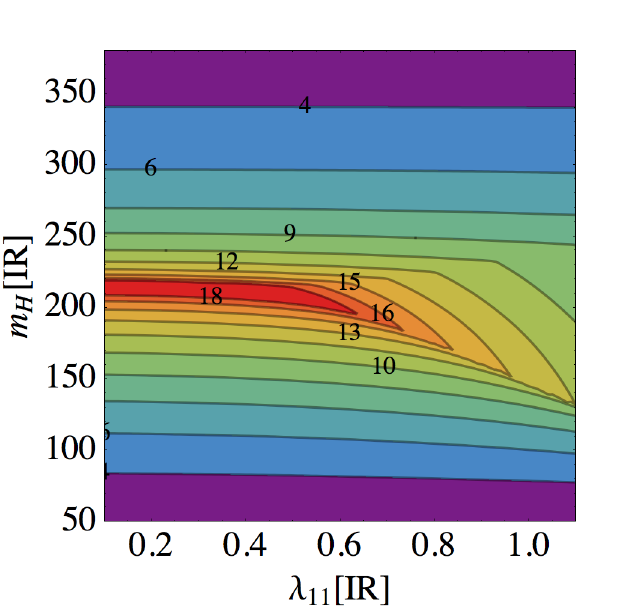}
\caption{\label{darkdolphin} We plot contours of fixed cutoff scale $\Lambda$, indicated by colors and numerical values in the plane spanned by $m_H$ and $\lambda_{11}$. The labels at the contours give $\mathrm{Log}_{10} (\Lambda/\mathrm{GeV})$. The dark scalar mass is fixed by the requirement that the scalar constitutes the complete dark matter relic density and the dark matter self coupling is $\lambda_{02}(k_{\rm IR}) = 10^{-2}$. None of the couplings exceeds the value 3.}
\end{figure}

In agreement with \cite{Gonderinger:2009jp}, we observe two main effects of dark matter. Both result from the bosonic nature of the dark matter scalar, which implies that $\lambda_{11}$ yields a positive contribution to the beta function for the Higgs self-coupling, cf.~app.~\ref{FRG}. Accordingly the dark matter scalar counteracts the effect of fermionic fluctuations. A larger $\lambda_{11}$ thus implies that a lower Higgs mass can be reached in the infrared while starting from the same UV cutoff scale, cf. fig.~\ref{CutoffVsMh}.
On the other hand, this positive contribution also enhances the divergence in $\lambda_{20}$ at larger values. Thus the region of larger Higgs masses is characterized by a faster approach to  the Landau pole and therefore a lower cutoff scale. Thus both the upper as well as the lower bound on the Higgs mass for a given cutoff scale $\Lambda$ move down. Both boundaries meet at a finite value of $\lambda_{11}$, see fig.~\ref{darkdolphin}.

Larger values of $\lambda_{02} (k_{\rm IR})$ result in a quicker approach to the Landau pole and are thus not conducive to a high cutoff scale $\Lambda$. It turns out that values $\lambda_{02}(k_{\rm IR})\leq10^{-2}$ do not result in a different shape of the $\Lambda$ contours.

Our functional RG approach includes nonperturbative threshold effects and accounts for spontaneous symmetry breaking during the RG flow. This allows to dynamically generate the Higgs mass during the RG flow. For a detailed study of threshold effects, see app.~\ref{thresholds}.

Let us stress that a main difference between our model and the Standard Model lies in the location of the region where $\Lambda \approx 10^{19 }\, \rm GeV$. Whereas it lies above 200 GeV in our case, the corresponding region in the Standard Model lies only a few GeV above the measured value of the Higgs mass \cite{Bezrukov:2012sa, Gonderinger:2009jp}. Accordingly, whereas approximately 100 GeV have to be bridged in our model, a much smaller difference of only a few GeV must be bridged in the Standard Model case.

\begin{figure}[!here]
\includegraphics[width=\linewidth]{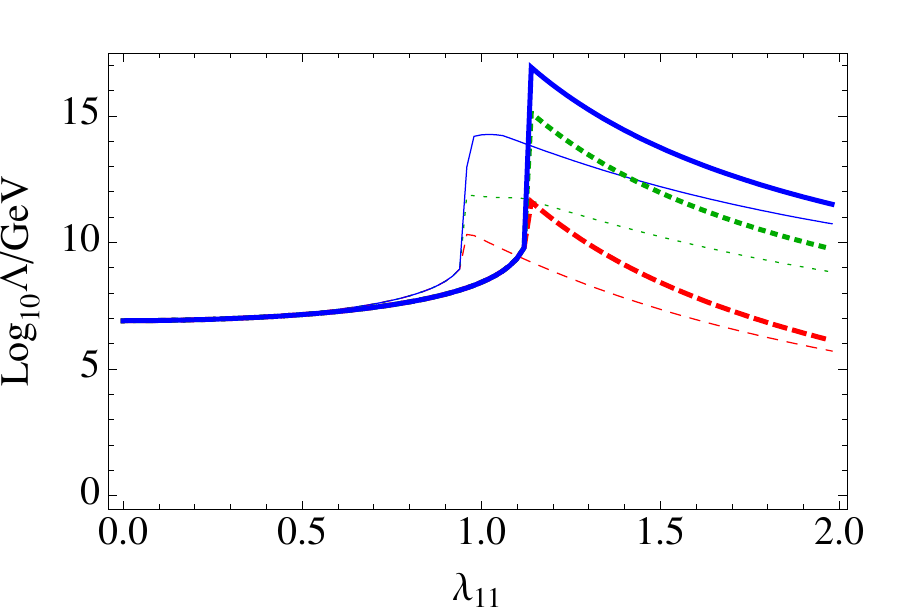}
\caption{\label{CutoffVsLambda11} We plot the UV cutoff $\Lambda$ as a function of the infrared value of $\lambda_{11}$ for $m_{H}= 126 \, \rm GeV$. Thick lines correspond to an infrared value of $\lambda_{02}= 10^{-3}$, and thin lines $\lambda_{02}= 0.7$. Blue continuous lines are for a maximal value of 100 for all couplings, green dotted for 10, and red dashed for 3.}
\end{figure}

Focusing on the physical value of the Higgs mass, $m_{H}= 126 \, \rm GeV$, we then evaluate the cutoff scale as a function of $\lambda_{11}$ and $\lambda_{02}$. We also test, how much the upper bound depends on the maximal allowed value of the couplings, cf. fig.~\ref{CutoffVsLambda11}. Up to $\lambda_{11} \leq 1$, the UV cutoff $\Lambda$ does not depend on the maximal allowed value for the couplings, as the cutoff is induced by the stability bound. 
For larger $\lambda_{11}$, the cutoff depends on the maximal value of $\lambda_{ij}$, as the cutoff scale is reached when at least one of the couplings increasing beyond the maximal value. A larger value of $\lambda_{02}$ implies an earlier onset of the triviality bound, and leads to smaller cutoff values at $\lambda_{11} \gtrsim 1$. Accordingly, a smaller $\lambda_{02}$ allows to reach higher cutoff scales, cf. thick lines in fig.~\ref{CutoffVsLambda11}.

Our analysis implies that the maximal UV cutoff for physical values of the Higgs and top mass that can be reached in this model is higher than without dark matter, but still lies significantly below the Planck scale. As our functional RG approach allows for the couplings to leave the perturbative regime, we can study a region of couplings beyond those allowed in the perturbative study in \cite{Gonderinger:2009jp}. We conclude that cutoff values at the GUT scale, i.e., $\Lambda=10^{16}\,\rm GeV$, could be reached. Still, the Planck scale seems to remain out of reach in this setting. Let us emphasize that these scales cannot be translated directly into values for the case of the Standard Model.

\subsection{The effect of dark matter -- new couplings}

We will now take into account the effect of higher order couplings in the effective potential.
As we discussed above, effective-field-theory arguments suggest that such couplings can easily be $\mathcal{O}(1)$  in units of the cutoff scale, and therefore play a significant role at high scales.

\begin{figure}[!here]
\includegraphics[width=\linewidth]{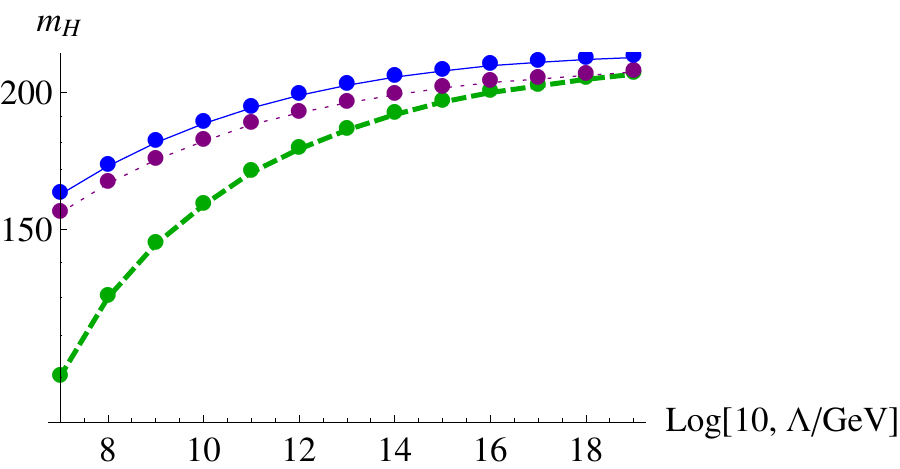}
\caption{\label{MhvsLambda} We plot the lower bound on the Higgs mass in GeV as a function of UV cutoff $\Lambda$ for the model without dark matter (blue thin line), including dark matter with $\lambda_{11}(k_{\rm IR})=0.5$ (purple dotted line). The green dashed line shows the values for the Higgs mass that we obtain when we include higher-order couplings, and start with $\lambda_{30}(k_{\rm UV})=3= \lambda_{03}$, $\lambda_{02}(k_{\rm UV})=0.5 $ and $\lambda_{20}(k_{\rm UV})= -0.25$. The UV value of the Yukawa-coupling is fixed to give the correct value of the top quark mass. Here, we neglect non-perturbative threshold effects.}
\end{figure}

We point out that given an ultraviolet cutoff scale $\Lambda$, it is possible to reach values for the Higgs mass that lie below the lower bound that arises in the case where the potential is restricted to contain only quartic couplings, cf. fig.~\ref{MhvsLambda}. 
Most importantly, the lower (green dashed) curve in fig.~\ref{MhvsLambda} is \emph{not} a lower bound on the Higgs mass, as even lower values can be obtained for the same choice of the cutoff, if a lower $\lambda_{20}(k_{\rm UV})$ is chosen. The values for the Higgs mass on that curve are predictions for the Higgs mass that arise from a given choice of values for the couplings in the ultraviolet.

Our most important conclusion regards the infrared value of the Higgs mass in the case of a cutoff at the Planck scale, $\Lambda \approx 10^{19}\, \rm GeV$. The combined effect of the dark matter scalar and the inclusion of higher order couplings leads to a value that is approximately 7 GeV below the lower bound arising in a quartic potential without dark matter (for the choice of UV initial conditions used in fig.~\ref{MhvsLambda}). We stress that a different choice of UV initial conditions could even yield masses below that value. Our model does not contain all Standard Model degrees of freedom, but we can still embark on a speculation for that case: Assuming that the combined effect of dark matter and generalized UV initial conditions yields a similar change in the infrared value of the Higgs mass, we conjecture that the Standard Model could be valid all the way up to the Planck scale.

Employing the beta functions including nonperturbative threshold effects is technically slightly more involved, as the UV initial conditions for the Higgs vev and the dark matter mass need to be fine-tuned. As can be seen in tab.~\ref{tab}, explicit integrations confirm the results presented in fig.~\ref{MhvsLambda}: The inclusion of dark matter and the generalization of the UV initial conditions leads to a significant decrease in the infrared value of the Higgs mass. 

The comparison of fig.~\ref{MhvsLambda} and tab.~\ref{tab} confirms that the qualitative side of this mechanism is independent of nonperturbative threshold effects, and thus applicable also within a setting where the Standard Model remains perturbative, i.e., the couplings remain bounded from above by, e.g., 1, all the way to the Planck scale.

\begin{table}[!t]
\caption{Higgs mass for different scenarios calculated with the FRG $\beta$ functions including threshold effects. We show the lower mass bounds in the case without higher order couplings, i.e. $\mathcal{O}(\phi^4)$, and a mass with higher order couplings, $\mathcal{O}(\phi^6)$ lying below the naive bound.
Here, $vev=246$ GeV and $m_{\mathrm{top}}=173$ GeV and when dark matter is included, the relic density relation is fulfilled with very good accuracy (deviation of $< 2\%$ in the dark matter mass). All higher couplings to order $\mathcal{O}(\phi^6)$ in the scalar fields $h$ and $S$ are included and we choose a stable UV potential with $\lambda_{20}(k_{\mathrm{UV}})=-0.1$ and $\lambda_{30}(k_{\mathrm{UV}})=\lambda_{03}(k_{\mathrm{UV}})=3$ and further $\lambda_{02}(k_{\mathrm{UV}})=0.01, \lambda_{21}(k_{\mathrm{UV}})= \lambda_{12}(k_{\mathrm{UV}})=0$. \label{tab}}
\begin{ruledtabular}\begin{tabular}{ccc}
$\Lambda_{\mathrm{UV}}=10^8 \mathrm{GeV} $ & $\lambda_{11}(k_{\mathrm{IR}})$ & $m_H/\mathrm{GeV}$\\ \hline
$\mathcal{O}(\phi^4)$, no dark matter &  0 & 141.1 \\
$\mathcal{O}(\phi^4)$ & 0.87  &  127.8 \\
$\mathcal{O}(\phi^6)$ & 0.87  &  111.1 \\ \\ \hline
$\Lambda_{\mathrm{UV}}=10^{10} \mathrm{GeV} $ & $\lambda_{11}(k_{\mathrm{IR}})$ & $m_H/\mathrm{GeV}$\\ \hline
$\mathcal{O}(\phi^4)$, no dark matter &  0 & 163.7 \\
$\mathcal{O}(\phi^4)$ & 0.77  &  150.7 \\
$\mathcal{O}(\phi^6)$ & 0.77  &  139.9 \\
\end{tabular}\end{ruledtabular}
\end{table}

The scale-dependence of the couplings shows the expected behavior: At the UV scale, $\lambda_{20}$ is negative, which would imply an instability if no higher-order couplings were present and thus rule out this particular microscopic theory. As $\lambda_{30}>0$, the potential is stable at the cutoff scale $\Lambda$. Towards lower scales, $\lambda_{30}$ quickly becomes very small, as expected. 

\begin{figure}[!t]
\includegraphics[width=0.8\linewidth]{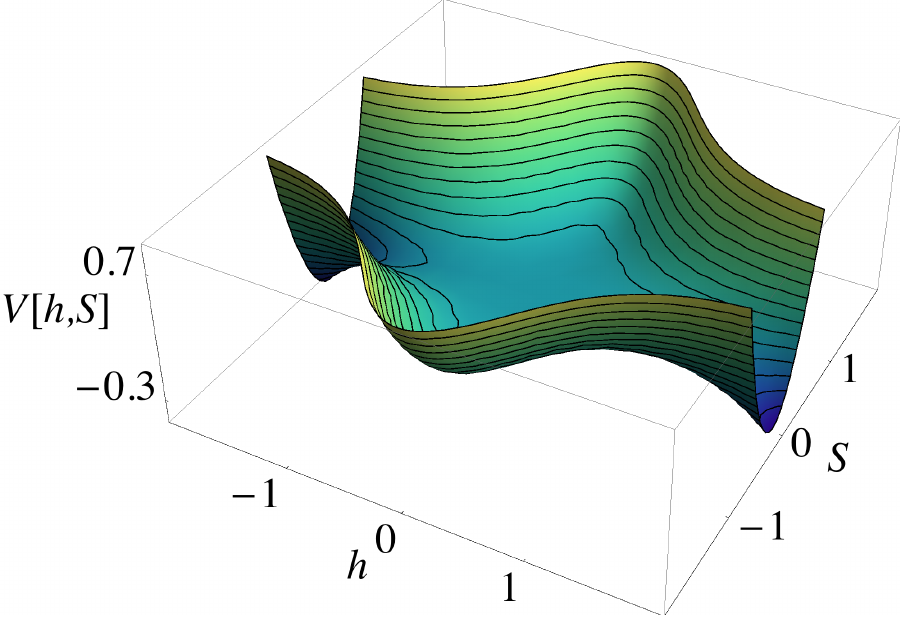}\\
\includegraphics[width=0.8\linewidth]{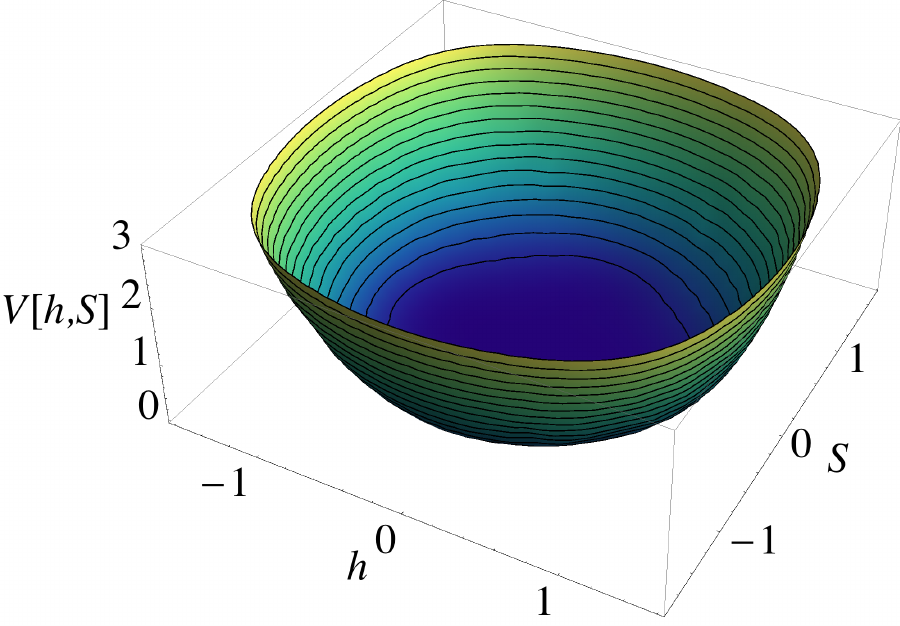}
\caption{\label{Potplots}We plot the potential at the UV cutoff scale $\Lambda$. Setting higher-order couplings to zero (upper panel) leads to an unstable bare potential. Including the couplings $\lambda_{30}$ and $\lambda_{03}$ (see above for values) leads to a stable bare potential  with its minimum at a vanishing value for the Higgs field.}
\end{figure}

Note that in our analysis 
we neglect couplings of higher order, such as $h^8$ couplings. Even if set to zero at the UV cutoff scale, these can be generated by the RG flow close to the UV scale. Towards the infrared, they run to zero quickly, and will not have a significant effect on the RG flow of the other couplings. A corresponding convergence study has been performed in \cite{Gies:2013fua}. Similar results are assumed to hold in the case of the Higgs portal.

Crucially, the potential in the UV is completely stable with its minimum lying at a vanishing value of the Higgs field, cf. fig.~\ref{Potplots}. A nonvanishing minimum, corresponding to $vev=246\, \rm GeV$ is generated during the RG flow towards the infrared.

Let us briefly discuss our choices for $\lambda_{11}$ and $m_S$ in the context of experimental searches for dark matter: Regions of parameter space, where $\lambda_{11} \sim 1$, are experimentally viable, while a significant part of the dark matter parameter space at low values of $\lambda_{11}$ has already been excluded, cf. \cite{Cline:2013gha}. The region that is of particular interest in the context of a low value of the Higgs mass therefore still remains viable. 
Future updates of the XENON experiment \cite{Aprile:2012nq, Aprile:2012zx,Beltrame:2013bba}, as well as the LUX experiment \cite{Fiorucci:2013yw,Woods:2013ufa}, will start to probe this interesting regime.

\section{Conclusions}

In our work, we address two major challenges faced by the Standard Model: Firstly, it does not contain an experimentally viable candidate for cold dark matter, which
is required to explain astrophysical and cosmological data in the absence of modifications of the gravitational dynamics. Secondly, many studies indicate that the measured value of the Higgs mass precludes an extension of the Standard Model all the way to the Planck scale. Here we use the simple observation that additional bosonic degrees of freedom, coupled to the Higgs via a quartic coupling, can contribute to a lowering of the bound on the infrared value of the Higgs mass,
as they yield a positive contribution to the $\beta$ function of the Higgs self-coupling.
At the same time, this model is highly attractive, as the additional scalar provides a stable dark matter candidate that can explain the complete dark matter relic density with only a simple additional scalar degree of freedom.
If successful, our program implies that the understanding of fundamental particles and their interactions spans a huge range of scales and physical phenomena. The main missing ingredient for a complete picture is then the inclusion of quantum gravitational dynamics beyond the Planck scale. This scenario is suggested by the experimental data coming from the LHC, implying that the value of the Higgs mass lies in a regime where an extension of range of validity of the Standard Model plus a dark matter scalar all the way up to the Planck scale could be possible.

We confirm previous studies that show that the inclusion of the additional scalar increases the UV cutoff scale of the Standard Model to higher scales. Here we point out that these studies were rather restricted in the following sense:
As the Standard Model is a low-energy effective field theory, no higher-order couplings are present at low energies. Simultaneously, canonical power counting arguments suggest that terms such as $h^6$  can have couplings of $\mathcal{O}(1)$ at UV scales. An analysis of the RG flow, bridging the gap to those high scales, must take these couplings into account. To our best knowledge, our study is the first to consider these couplings in the context of the Higgs dark matter portal. We point out that earlier studies relied on choosing very specific initial conditions for the RG flow at high scales, where all higher order couplings are set to zero. As no symmetry-principle forbids their presence, one should generically expect that an ultraviolet completion of the Standard Model will induce these couplings at the UV scale.

We employ functional Renormalization Group tools to analyze the RG flow in a toy model of the Higgs sector of the Standard Model plus dark matter, including two scalars (Higgs and dark matter scalar), and a fermion (top quark). 
We provide a proof of principle, that the inclusion of higher-order couplings in this system  
 allows us to obtain the measured value of the Higgs and top mass in the infrared, while starting from a UV-stable potential at higher cutoff scales $\Lambda$. Simultaneously, the dark matter scalar accounts for the complete dark matter relic density in a viable region of the parameter space of that model. As we have shown, the inclusion of dark matter and the generalization of the UV bare potential allows to reach Higgs masses lying several GeV below the lower bound from a quartic potential, when the cutoff is chosen to lie at the Planck scale.
Since we employ a toy-model, extrapolations of the numerical results to the case of the Standard Model have to be taken with a grain of salt, but could actually imply the possibility of a cutoff scale that is as high the Planck scale. As the lower bound on the Higgs mass in the case of the Standard Model with a cutoff at the Planck scale lies at $m_H =129 \, \rm GeV$, it is very tempting to speculate that the inclusion of dark matter and higher-order couplings will provide the additional $\sim 4\, \rm GeV$ necessary to accommodate the measured value of the Higgs mass.

 A significant part of the dark-matter parameter space that seems to be of interest for this scenario will be probed in the near future, e.g., by the XENON experiment.
We emphasize that it is a combination of two effects that could allow the Standard Model to bridge the way from low scales to the Planck scale: Firstly, the inclusion of a dark matter scalar extends the UV cutoff scale to higher values. In a second step, higher-order couplings can play a role on scales below the UV cutoff scale. This could provide an ultraviolet stable potential at the Planck scale. The RG flow then forces the higher-order couplings to vanish, while the quartic couplings, which are negative at the cutoff scale, grow to reach positive values. Going further in the direction of the infrared, the RG flow is then essentially indistinguishable from a flow including only quartic couplings.

A study of the convergence of the polynomial expansion of the effective potential during the RG flow is beyond the scope of this analysis. We defer this important point to future work, and only point out here that a numerical analysis of the RG flow of the effective potential not relying on a polynomial expansion is better suited to understand all properties -- global and local -- of the effective potential at all scales. 
This will allow us to conduct a comprehensive study of the space of bare potentials as well as an extensive analysis of the dark matter parameter space.
In the future, more detailed studies of the proposed scenario, including all relevant Standard Model degrees of freedom, will allow to provide a quantitatively precise determination of possible UV stable potentials and the corresponding cutoff scales.

{\emph{Acknowledgements:} 
We are indebted to H.~Gies for providing important comments on a first version of this manuscript.
We would  also like to thank J.~J\"ackel for helpful comments on the manuscript, H.~Gies and R.~Sondenheimer for helpful correspondence and T.~Plehn for interesting discussions. A.~E. would also like to thank B.~D\"obrich for helpful discussion on dark matter searches.
A.~E. gratefully acknowledges the hospitality of the Institute for Theoretical Physics at the Ruprecht-Karls-University of Heidelberg during the final stages of this project.
This research was supported in part by Perimeter Institute for Theoretical Physics. Research at Perimeter Institute is supported by the Government of Canada through Industry Canada and by the Province of Ontario through the Ministry of Research and Innovation. M.~M.~S. is supported by the grant ERC- AdG-290623.}

\begin{appendix}
\section{Derivation of $\beta$ functions}\label{FRG}

We employ the functional Renormalization Group to extract the $\beta$ functions of running couplings. The $\beta$ functions are obtained from a momentum-shell wise integration over the quantum fluctuations in the path integral $Z[J] =\mathrm{exp}(W[J])= \int_\Lambda \mathcal{D} \varphi e^{- S[\varphi] + J \cdot \varphi}$ with the microscopic action $S[\varphi]$ and the source $J$, as suggested by Wilson. This is achieved by introducing a momentum-scale dependent mass-term of the form $\Delta S_k[\varphi]=\int d^dp\, \varphi(-p) \cdot R_k (p) \cdot \varphi(p)$ into the action in the path integral, where
\bea
R_k(p) &=& 0, \quad \mbox{ for } p^2>k^2,\nonumber\\
R_k(p)&\geq& 0, \quad \mbox{ for } p^2 <k^2, \nonumber\\
R_k(p) &\rightarrow& 0, \quad \mbox{ for } k^2 \rightarrow 0.
\eea
The first condition ensures that for high-momentum modes, the path-integral is unmodified and they are integrated out, contributing to the effective action for low-momentum modes. The second condition results in a suppression of low-momentum modes which do not contribute to the effective dynamics. Finally the last condition implies that the standard effective action is reached in the limit $k \rightarrow 0$.
A particular function that satisfies these requirements and that we will use in the following is given by
\bea
R_{k\, b} (p) &=& (k^2-p^2) \theta(k^2-p^2), \nonumber\\
 R_{k\, f}(p) &=&\slashed{p} \left(\sqrt{\frac{k^2}{p^2}}-1 \right)\theta(k^2-p^2),
\eea
where the first line holds for scalar bosonic, and the second for fermionic degrees of freedom.

We obtain the field expectation value from the relation $\phi=\langle \varphi \rangle=Z^{-1}(\delta Z/\delta J)$.
It is then possible to define a scale-dependent effective action $\Gamma_k$ for the momentum-modes below and at $k$ by
\be
\Gamma_k = \mathrm{sup}_J\left( J\cdot \phi - W[J]\right)-\Delta S_k[\phi]
\ee
Integrating out quantum fluctuations in the momentum shell $\delta k$ results in a change of the couplings $g_i(k)$ in the effective action $\Gamma_k=\sum_i g_i(k)\mathcal{O}_i(\phi)$ where $\mathcal{O}_i(\phi)$ are local operators compatible with the symmetries of the model. The regularization prescription given above provides an interpolating trajectory for $\Gamma_k$ in theory space starting with the classical action $S$ at the UV scale $\Lambda$ and yielding the full quantum effective action $\Gamma$ at $k=0$. We will ultimately be interested in an extraction of the corresponding $\beta$ functions for the couplings and thus require an equation giving the scale dependence of the effective action. The Wetterich equation \cite{Wetterich:1992yh} can be derived exactly and reads
\be
\partial_t \Gamma_k=\frac{1}{2} {\rm STr} \left( \Gamma_k^{(2)}+R_k\right)^{-1} \partial_t R_k.
\ee
Herein $t = \ln k$ and $\Gamma_k^{(2)} = \frac{\overset{\rightarrow}\delta}{\delta \phi_a (-p)} \Gamma_k  \frac{\overset{\leftarrow}\delta}{\delta \phi_b (q)}$. $\phi_a$ is a superfield that collects all bosonic and fermionic degrees of freedom of the theory. In our case it is given by $\phi_a (p) = (h(p), S(p), \psi(p),\bar\psi^T(-p))$.
The supertrace $\rm STr$ is a summation over all fields and includes a negative sign for fermionic degrees of freedom. Further, the trace includes a momentum-integral, giving the Wetterich equation an effective one-loop form. A summation over all internal indices and spacetime indices is also implied.

The Wetterich equation is highly useful as it is structurally a one-loop equation, which encodes nonperturbative effects through the use of the full nonperturbative propagator. Its derivation does not rely on any assumption, e.g., about the smallness of couplings, and it therefore holds exactly. For practical purposes approximations are necessary, though. The main reason lies in the fact that quantum fluctuations generate all operators which are compatible with the symmetries of the theory, which usually amount to infinitely many. The effective action at a scale $k$ is thus given by an infinite sum of terms. The Wetterich equation then decomposes into an infinite tower of coupled differential equations for the couplings of these operators. This structure is highly similar to that of Dyson-Schwinger equations for the $n$ point functions of a theory. In practice a typically finite ansatz for the effective action is necessary in order to extract results from the Wetterich equation.
For reviews of this method see \cite{Berges:2000ew,Polonyi:2001se,Pawlowski:2005xe,Gies:2006wv,Delamotte:2007pf,Rosten:2010vm,metzner2011}.

We will follow the same route here and choose the following truncation
\bea
\Gamma_k &=& \int d^4 x \Big\{\bar\psi i \slashed{\partial} \psi
+\frac{1}{2}(\partial_\mu h)^2+\frac{1}{2}(\partial_\mu S)^2\nonumber\\
&&\quad\quad+i \bar y(k) h\bar\psi\psi+\bar V_k(h,S)\Big\}\,,
\eea
with an effective potential of the form
\bea
\bar V_{k}(h, S)&=&  \frac{\bar m_h(k)^2}{2} h^2+\bar m_S(k)^2 S^2\nonumber\\
&&+\frac{\bar\lambda_{20}(k)}{8}h^4+\frac{\bar\lambda_{02}(k)}{8}S^4+\frac{\bar\lambda_{11}(k)}{4}h^2S^2\nonumber\\
&&+\frac{\bar\lambda_{30}(k)}{48}h^6+\frac{\bar\lambda_{03}(k)}{48}S^6+...
\eea
As we are considering the RG flow of a theory with spontaneous symmetry breaking, we need to adapt our truncation to that situation. To correctly account for the effects of quantum fluctuations in the symmetry-broken regime, we use the following ansatz for the effective potential
\bea\label{eq:ssb}
\bar V_{k\, \rm SB}(h, S)&=& \frac{\bar m_S(k)^2}{2} S^2+\frac{\bar\lambda_{20}(k)}{2}\left(\frac{h^2}{2} - \bar\kappa_{h}\right)^2\\
&+&\frac{\bar\lambda_{02}(k)}{8}S^4+\frac{\bar\lambda_{11}(k)}{2}\left(\frac{h^2}{2} - \bar\kappa_{h}\right)S^2\nonumber\\
&+&\frac{\bar\lambda_{30}(k)}{48}\left(\frac{h^2}{2} - \bar\kappa_{h}\right)^3+\frac{\bar\lambda_{03}(k)}{48}S^6+...\nonumber
\eea
Generically the RG flow for this class of theories shows the following behavior: Starting in the symmetric regime at high scales, quantum fluctuations lower the Higgs mass, until it reaches zero. At this point, the symmetry is broken spontaneously and we switch our parameterization of the effective action to the second parameterization. This ensures that our expansion point for the effective potential always corresponds to the minimum of the effective potential, and not a saddle-point.

We introduce the invariants $\rho_h = \frac{h^2}{2}$ and $\rho_S = \frac{S^2}{2}$ and then make a transition to dimensionless couplings by defining:
\bea
\tilde\rho_{h,S}=k^{-2}\rho_{h,S},\ y(k)=\bar y(k),\ u_k=k^{-4}\bar V_k
\eea
Note that by introducing $\rho_h$ and $\rho_S$ the dimensionless effective potential is a function of their dimensionless versions $u_k=u_k(\tilde\rho_h,\tilde\rho_S)$.
The RG flow can be studied both for the dimensionless as well as the dimensionful couplings. The only difference is, that the former includes a ``classical" running, that is associated to a rescaling of a classical dimensionful quantity under a change of units. The RG flow can easily be translated from one to the other formulation.

Using this truncation, we derive the following $\beta$ functions for the Yukawa coupling and for the dimensionless effective potential. In the symmetric regime, we find
\bea
\beta_{y\, \rm SYM} &=&\frac{1}{8 \pi^2} y^4 \frac{2+m_h^2}{\left(1+m_h^2 \right)^2},\nonumber\\
\eea

whereas the form in the symmetry-broken regime reads
\bea
\beta_{y\, \rm SB}&=&\frac{1}{4 \pi^2} y^4 \frac{1}{(1+2 y^2 \kappa_h)^3 (1+2 \kappa_h \lambda_{20})^3} \\
&{}&- \frac{1}{2 \pi^2} y^4 \kappa_{h} \frac{(3+2 \kappa_h \lambda_{20}) (\lambda_{20}+ \kappa_h \lambda_{30})}{(1+2 y^2 \kappa_h)^3 (1+2 \kappa_h \lambda_{20})^3} \nonumber\\
&{}& - \frac{1}{2 \pi^2} y^8 \kappa_h^2 \frac{1+14 \kappa_h \lambda_{20} + 8 \kappa_h^2 \lambda_{30}}{(1+2 y^2 \kappa_h)^3 (1+2 \kappa_h \lambda_{20})^3} \nonumber\\
&{}&- \frac{y^6}{4 \pi^2} \kappa_h \frac{3+6 \kappa_h \lambda_{20}(7+4 \kappa_h \lambda_{20})}{(1+2 y^2 \kappa_h)^3 (1+2 \kappa_h \lambda_{20})^3}\nonumber\\
&{}&- \frac{y^6}{4 \pi^2} \kappa_h \frac{4 \kappa_h^2 (5+2 \kappa_h \lambda_{20})\lambda_{30}}{(1+2 y^2 \kappa_h)^3 (1+2 \kappa_h \lambda_{20})^3}.\nonumber
\eea
For the effective potential, we obtain the following form:
\bea
\partial_t u_k \hspace{-0.15cm}&=&\hspace{-0.15cm}-4 u_k+2\tilde\rho_h u_k^{(1,0)} + 2\tilde\rho_S u_k^{(0,1)}\nonumber\\
	 &&+I_{R,h}^d(\omega_S,\omega_h,\omega_{h S})+I_{R,S}^d(\omega_h,\omega_S,\omega_{h S})\nonumber\\
&& + \partial_t u_k\vert_{\psi}\,.\label{potflow}
\eea
Herein, the first line arises from canonical dimensionality. The $u_k^{(1,0)}$ and $u_k^{(0,1)}$ denote the derivatives with respect to the first and second argument of $u_k$, respectively. The subsequent line corresponds to the non-perturbative loop contributions of the two fields $h$ and $S$.  The fermionic loop contribution is given by
\be
\partial_t u_k\vert_{\psi}= -\frac{1}{8 \pi^2} \frac{1}{1+2 \tilde\rho_h y^2}.
\ee
We have defined the threshold function,
\be
I_{R,i}^{d}(x,y,z)=\frac{4v_d}{d}\frac{1+x}{(1+x)(1+y)-z},
\ee
with the volume element $v_d^{-1}=2^{d+1}\pi^{d/2}\Gamma(\frac{d}{2})$, where $d=4$ in our case. The arguments in the flow equation \eqref{potflow} read
\bea
\omega_h&=&u_k^{(1,0)}+2\tilde\rho_h u_k^{(2,0)},\\
\omega_S&=&u_k^{(0,1)}+2\tilde\rho_S u_k^{(0,2)},\\
\omega_{hS}&=&4\tilde\rho_h \tilde\rho_S \big(u_k^{(1,1)}\big)^2.
\eea

\eqref{potflow} summarizes the $\beta$ functions for all couplings in the effective potential in a very compact form. For our explicit evaluation of the RG trajectories, we insert our ansatz for the effective potential into that equation and can then project on the appropriate powers of $\tilde\rho_h$ and $\tilde\rho_S$ to obtain the corresponding $\beta$ functions.
Specifically, we use 
\be
\partial_t \lambda_{l,m}=\label{eq:projections2}\!\left(\partial_t u_k^{(l,m)}\!+\!u_k^{(l+1,m)}\partial_t\kappa_h\right)\!\Big|_{\tilde\rho_h=\kappa_h \atop{\tilde\rho_S=0}}\!.
\ee
To distinguish between the symmetric and the symmetry-broken regime, it is simply necessary to use $\kappa_h =0$ and $\kappa_h \neq 0$.
To obtain $\beta_{\kappa_h}$, we use the following projection prescription
\be
\beta_{\kappa_h} = -\frac{\partial_t u_k^{(1,0)}}{u_k^{(2,0)}}\Big|_{\tilde\rho_h=\kappa_h \atop{\tilde\rho_S=0}}
\ee

As a concrete example, the $\beta$ function for $\lambda_{11}$ in the symmetric regime reads
\bea
\beta_{\lambda_{11}} = \partial_t \lambda_{11} &=& \frac{3 \lambda_{02} \lambda_{11}}{16 \pi^2 (1+m_S^2)^3} + \frac{3 \lambda_{11} \lambda_{20}}{ 16 \pi^2 (1+m_h^2)^3}\nonumber\\
&{}& + \frac{\lambda_{11}^2}{ 8 \pi^2} \frac{(2+m_S^2 +m_h^2)}{(1+m_S^2)^2(1+m_h^2)^2}.
\eea

In the appropriate limit, where the interaction with the dark matter scalar vanishes, our $\beta$ functions reduce to those of \cite{Gies:2009hq}.

\section{Threshold effects on the cutoff scale}\label{thresholds}
\begin{figure}[!here]
\includegraphics[width=0.8\linewidth]{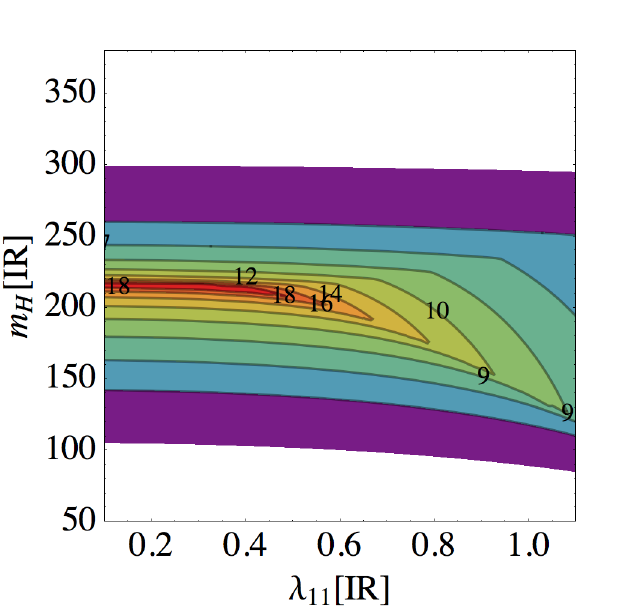}
\caption{\label{darkdolphin_pert} We plot contours of fixed cutoff scale $\Lambda$, indicated by colors and numerical values in the plane spanned by $m_H$ and $\lambda_{11}$. The dark scalar mass is fixed by the requirement that the scalar constitutes the complete dark matter relic density. The dark matter self coupling $\lambda_{02} = 10^{-2}$. None of the couplings exceeds the value 3.}
\end{figure}

The functional Renormalization Group framework allows for an automatic (partial) resummation of perturbative effects through a nonperturbative propagator, that enters the beta functions in the form $(1+m_{\Phi}^2)^{-1}$, where $\Phi= \{h, S, \psi \}$. In the symmetry-broken regime, the corresponding masses are proportional to the Higgs self-interaction in the case of $m_h$, and to the Yukawa coupling in the case of $m_{\psi}$. In the following we study the effect of these nonperturbative thresholds on the cutoff scale $\Lambda$. We accordingly neglect these effects on the RG flow, and evaluate the cutoff scale for purely perturbative $\beta$ functions.
Fig.~\ref{darkdolphin_pert}  in comparison to fig.~\ref{darkdolphin} clearly reveals that the band where the highest cutoff scales can be reached, is significantly slimmer. In particular, the region of high Higgs masses exhibits significantly lower values of the UV cutoff scale. This is immediately plausible, as denominators of the form $(1+ 2 \kappa_{h} \lambda_{20})^{-1}$ clearly lead to a suppression of the corresponding terms in the $\beta$ function, when $\lambda_{20}$ reaches large values. Accordingly, nonperturbative threshold effects slow the approach towards the Landau pole. Thereby the same value of the Higgs mass leads to a lower UV cutoff scale, when these effects are ignored.

\end{appendix}

\end{document}